\documentclass[conference, letterpaper]{IEEEtran}

\newcommand{\route}{R}

\newcommand{\nowp}{k}
\newcommand{\tw}{\texttt{tw}}

\newcommand{\sepname}{s}
\newcommand{\bagname}{b}

\ifx\pdfoutput\undefined
\usepackage[hypertex]{hyperref}
\else
\usepackage[pdftex,hypertexnames=false]{hyperref}
\fi

\usepackage{url}
\usepackage{microtype}

\usepackage{etoolbox}

\usepackage{amsmath,amssymb,array,color,colortbl,graphicx,multirow,mathtools}
\usepackage{microtype}
\usepackage{comment}
\usepackage{todonotes}
\usepackage{balance}
\usepackage{cleveref}
\usepackage{threeparttable}
\usepackage{graphicx}
\usepackage{comment}
\usepackage{listings,color}
\usepackage{verbatim}
\usepackage{multirow}

\usepackage{listings}

\usepackage{framed,color}

\definecolor{mygreen}{rgb}{0,0.5,0}

\usepackage[para]{footmisc}
\newtheorem{theorem}{Theorem}
\newtheorem{proposition}{Proposition}

\newtheorem{corollary}{Corollary}

\newtheorem{lemma}{Lemma}

\newtheorem{observation}[proposition]{Observation}

\newenvironment{Claim}[1]{\par\noindent\underline{Claim:}\space#1}{}
\newenvironment{ClaimProof}[1]{\par\noindent\underline{Proof:}\space#1}{\hfill~$\square$}

\crefname{claim}{claim}{claims}
\crefname{observation}{observation}{observations}
\crefname{item}{item}{items}
\crefname{case}{case}{cases}

\usepackage{xspace}
\newcommand{\WRP}{WRP\xspace}

\usetikzlibrary{calc}
\usetikzlibrary{chains}
\usetikzlibrary{patterns}
\usetikzlibrary{decorations.pathreplacing}
\usetikzlibrary{er}
\usetikzlibrary{shapes}
\usetikzlibrary{shapes.multipart}
\usetikzlibrary{arrows}
\tikzset{markovstate/.style={shape=circle,draw=black,align=center,inner sep=0,minimum width=7mm}}
\tikzset{markovedge/.style={-latex}}
\tikzset{label/.style={}}
\tikzset{pkt/.style={draw,rectangle,fill=gray!15,minimum height=25pt,align=center,font=\footnotesize}}
\tikzset{pktlen/.style={above=-2pt,font=\scriptsize}}

\definecolor{mycolor}{RGB}{0,200,100}
\begin{document}
\title{{\LARGE Walk the Line:}\\Charting the Complexity Landscape\\ of Waypoint Routing}

\title{Charting the Complexity Landscape\\ of Waypoint Routing}

\title{These NFVs Are Made for Walkin'}

\title{These VNFs Are Made for Walking}

\title{Charting the Complexity Landscape\\of Waypoint Routing}

%%%%%%%%%%%%%%%%%%%%%%ICNP
%\author{Submission \# XYZ, \pageref*{lastpage} pages, including references}
%%%%%%%%%%%%%%%%%%%%%%ARXIV
\IEEEoverridecommandlockouts
\author{\IEEEauthorblockN{Saeed Akhoondian Amiri$^*$
\thanks{$^*$
Saeed Amiri's research partly supported by the European Research
Council (ERC) under the European Union’s Horizon 2020 research and innovation programme
(grant agreement No 648527).
}
}
\IEEEauthorblockA{TU Berlin, Germany~\\
\url{saeed.amiri@tu-berlin.de}~}
\and
\IEEEauthorblockN{Klaus-Tycho Foerster}
\IEEEauthorblockA{Aalborg University, Denmark\\
\url{ktfoerster@cs.aau.dk}}
\and
\IEEEauthorblockN{Riko Jacob}
\IEEEauthorblockA{IT University of Copenhagen, Denmark\\
\url{rikj@itu.dk}}
\and
\IEEEauthorblockN{Stefan Schmid}
\IEEEauthorblockA{Aalborg University, Denmark\\
\url{schmiste@cs.aau.dk}}}

\author{Saeed Akhoondian Amiri$^1$ \quad Klaus-Tycho Foerster$^2$  \quad Riko Jacob$^3$ \quad Stefan Schmid$^2$ \\
{\small $^1$ TU Berlin, Germany \quad $^2$ Aalborg University, Denmark \quad 
$^3$ IT University of Copenhagen, Denmark}
}

%\author{Submission 162\\ \pageref*{lastpage} pages} % in total}
%\author{Submission \#162, 10 pages}

%, including references}

\maketitle
\thispagestyle{plain}
\pagestyle{plain}

\begin{abstract}
Modern computer networks support interesting new routing models in which
traffic flows from a source~$s$ to
a destination~$t$ can be flexibly steered
through a sequence of \emph{waypoints}, such as (hardware) 
middleboxes or (virtualized) network functions (VNFs),
to create innovative network services like service chains or segment routing.
While the benefits and technological challenges of providing such routing models 
have been articulated and studied intensively over the last years, 
much less is known about the underlying
algorithmic traffic routing problems.
This paper shows that the waypoint routing problem
features a deep combinatorial structure, and
we establish interesting connections to 
several classic graph theoretical problems.
We find that the difficulty of the waypoint routing problem
depends on the specific setting, and 
chart a comprehensive landscape of the computational
complexity. In particular, we derive several NP-hardness results,
but we also demonstrate that exact
polynomial-time algorithms
exist for a wide range of practically relevant scenarios. 
\end{abstract}

\section{Introduction}\label{sec:intro}
\subsection{The Motivation: Service Chaining and Segment Routing}

We currently witness two trends
related to the increasing number 
of middleboxes (e.g., firewalls, proxies,
traffic optimizers, etc.) in computer networks  
(in the order of the number of routers~\cite{someone}):
First, there is a push towards virtualizing
middleboxes and network functions, 
enabling faster and more flexible deployments
(not only at the network edge),
and
 reducing costs. Second, 
over the last years, innovative new
network services have been promoted by industry and standardization
institutes~\cite{etsi}, by \emph{composing}
network functions to \emph{service chains}~\cite{merlin,unify,stefano-sigc}. 
The benefits and technological challenges of implementing 
such more complex network services have been studied intensively,
especially in the context of Software-Defined Networks (SDNs) and
Network Function Virtualization~(NFV), 
%which introduce unprecedented
introducing unprecedented
flexibilities on how traffic can be steered through flexibly allocated
(virtualized) network functions (VNFs).

However, much less is known today about the algorithmic
challenges underlying the routing through such 
middleboxes or network functions, henceforth simply
called \emph{waypoints}. 
In a nutshell, the underlying algorithmic problem 
is the following: How to route a flow 
(of a certain size) from a given source~$s$
to a destination~$t$, \emph{via a sequence of~$k$ waypoints} 
$(w_1,\ldots,w_k)$?
The allocated flow needs to respect capacity constraints,
and ideally, be as short as possible. 

The problem can come in many different
flavors, depending on whether a shortest or just a feasible route 
needs to be computed, depending on 
the number~$k$ of waypoints, depending on the type
of the underlying network (e.g., directed vs undirected,
Clos vs arbitrary topology), etc.
Moreover, as middleboxes provide different functionality
(mostly security and performance related), waypoints may
or may not be \emph{flow-conserving}: 
e.g., a tunnel entry point may \emph{increase} the
packet size (by adding an encapsulation header)
whereas a wide-area network optimizer may \emph{decrease} the
packet size (by compressing the packet).

The goal of this paper is to develop 
algorithmic techniques to solve the different variants of the waypoint
 routing problem, as well as to explore limitations due to computational intractability.

\subsection{The Problem: Waypoint Routing}

More formally, 
%the inputs to the \emph{waypoint routing problem} are:
inputs to the \emph{waypoint routing problem} are:
\begin{enumerate}
\item \textbf{A network:} represented as a graph~$G=(V,E)$,
where~$V$ is the set of~$n=|V|$ switches/routers/middleboxes (i.e.,~the nodes) and
where the set~$E$ of~$m=|E|$ links can either be undirected or directed,
depending on the scenario. Moreover, each link~$e\in E$ may have a bandwidth
capacity~$c(e)$ and weights $\omega(e)$ (describing costs), both non-negative. If not stated otherwise, we assume that~$c(e)=1$ and $\omega(e)=1$ for all~$e\in E$.
\item \textbf{A source-destination pair~$(s,t)$ and a sequence of
    waypoints~$(w_1,\ldots,w_k)$:}
which need to be traversed along the way from~$s$ to~$t$, forming
a route~$(s,w_1,\ldots,w_k,t)$.
Unless specified otherwise, we will assume at most one waypoint per node, though it may be that $s=t$.  
Waypoints may also change the traffic rate:
We will denote the demand from~$s$ to~$w_1$ by~$d_0$, from~$w_1$ to~$w_2$
by~$d_1$, etc. That said, if not stated explicitly otherwise, we will assume that~$d_0=d_1=\ldots=d_k=1$,
and refer to this scenario as \emph{flow-conserving}.
\end{enumerate}

In general, we are interested in shortest routes (an \textbf{optimization problem}),
i.e., routes of minimal length~$|R|$, such that link capacities
are respected. However, we also consider the \emph{feasibility} of such
routes: is it possible to route such a flow without violating 
link capacities at all (a \textbf{decision problem})?
Sometimes, minimizing the total route length alone may not be
enough, but additional, \textbf{hard constraints} on the distance (or stretch)
between a terminal and a waypoint or between waypoints may be imposed. 

\begin{table*}[t!]

\centering
\setlength\extrarowheight{2pt}
\begin{threeparttable}
%\resizebox{\linewidth}{!}{%
\begin{tabular}{>{\centering\arraybackslash}p{0.8in}>{\centering\arraybackslash}p{0.89in}>{\centering\arraybackslash}p{0.99in}>{\centering\arraybackslash}p{0.99in}>{\centering\arraybackslash}p{1.24in}>{\centering\arraybackslash}p{0.7in}}
&
  \multicolumn{1}{c|}{\textbf{$\#$ Waypoints}}
&
  \textbf{Feasible}
&
  \textbf{Optimal}
&
  \textbf{Demand Change Feasible }
&
  \textbf{Optimal}
\\ \hline\hline
\multirow{3}{*}{\textbf{Undirected}}& \multicolumn{1}{c|}{1} & \multicolumn{2}{c|}{\textbf{P} (Thm.~\ref{thm:easy-undirected-1wp})} & \multicolumn{2}{c}{\multirow{2}{*}{\textbf{\textbf{Strongly NPC}} (Thm.~\ref{thm:und-cap})}}\\ \cline{2-4}
& \multicolumn{1}{c|}{constant} & \multicolumn{1}{c|}{\textbf{P} (Thm.~\ref{thm:easy-undirected-kwp})}& \multicolumn{1}{c|}{?
%\tnote{$\ddagger$}
}  \\ \cline{2-6}
& \multicolumn{1}{c|}{arbitrary} & \multicolumn{4}{c} {\textbf{Strongly NPC} (Thm.~\ref{thm:euler-np-hard})} \\ \cline{1-6}
\multirow{3}{*}{\textbf{Directed}}& \multicolumn{1}{c|}{1} & \multicolumn{4}{c}{\multirow{3}{*}{\textbf{\textbf{Strongly NPC}} (Thm.~\ref{thm:hard-directed})}} \\ \cline{2-2}
& \multicolumn{1}{c|}{constant} \\ \cline{2-2}
& \multicolumn{1}{c|}{arbitrary} \\ \cline{1-6}
%{\textbf{P} (Thm.~\ref{thm:fptund})
\end{tabular}
\end{threeparttable}
\vspace{0.1cm}
\centering \caption{Overview of the Complexity Landscape for Waypoint Routing in general graphs. }
\label{tbl:small}
\end{table*}
\subsection{Novelty: It's a walk!}

\begin{figure}[h]
%\vspace{-0.3cm}
	\centering
	\begin{tikzpicture}[auto]
	\node [markovstate] (s) at (0, 0) {$s$};
	\node [markovstate] (t) at (2, 0) {$t$};
	\node [markovstate] (w) at (-2, 0) {$w$};
	
	\draw [markovedge, bend right=10] (s) to (t);
	\draw [markovedge, bend right=10] (t) to (s);
	\draw [markovedge, bend right=10] (s) to (w);
	\draw [markovedge, bend right=10] (w) to (s);
	\draw [markovedge, bend right=10] (w) to (s);
	
	\draw [markovedge, ultra thick, draw=red!200] plot [smooth, tension=0.4] coordinates {(-0.25,0.25) (-1.75,0.25)};
	\draw [markovedge, ultra thick, draw=blue!200] plot [smooth, tension=0.4] coordinates {(-1.75,-0.25) (1.75,-0.25)};
	
	\end{tikzpicture}
	%\vspace{-0.2cm}
	\caption{A route~$(s,w,t)$ in the depicted network must contain a loop.
	The only solution is the walk~$s,w,s,t$, resulting from concatenating the red~$(s,w)$ and blue~$(w,t)$ paths.
	 It can hence not be described as a simple path. 
	}
	\label{fig:introductory}
	%\vspace{-0.25cm}
\end{figure}
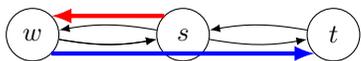

We will show that the waypoint routing problem is related to some 
classic and deep combinatorial problems, in particular the 
disjoint path
problem~\cite{thore-icalp,cygan2013planar,seymour1980disjoint12}
and the~$k$-cycle problem~\cite{thore-soda}.
In contrast to these problems, however, 
the basic waypoint routing problem considered in this paper comes with 
a fundamental twist: 
routes are not restricted to form simple paths, but can rather form 
arbitrary \emph{walks}, as long as capacity constraints in the underlying 
network are respected. Indeed, often feasible routes do not exist if restricted to 
a simple path, see Fig.~\ref{fig:introductory}
for an example in which any feasible route must contain a loop.

The problem is non-trivial. For example, consider the seemingly 
simple problem of routing via a \emph{single} waypoint, i.e., a
route of the form~$(s,w,t)$. 
A naive algorithm could try to first compute a shortest path from~$s$ to~$w$, 
deduct the resources consumed along this path, and finally compute a shortest 
path (subject to capacity constraints) from~$w$ to~$t$ on the remaining graph. 
However, as we will see shortly, such a greedy algorithm is doomed to fail;
rather, route segments between endpoints and waypoints 
must be \emph{jointly optimized}.

\subsection{Our Contributions}

This paper initiates the algorithmic study
of the waypoint routing problem underlying many
modern networking applications, such as service chaining~\cite{merlin}
(where traffic needs to be steered through network functions),
hybrid SDNs~\cite{panopticon} (where traffic is steered through OpenFlow
switches) or in segment routing~\cite{filsfils2015segment} 
(where MPLS labels are
updated at segment endpoints).

We show that whether and how efficiently
a \emph{feasible} or \emph{shortest} waypoint route can be found
depends on the scenario, and 
chart a complexity landscape 
of the waypoint routing problem, presenting a comprehensive set of 
NP-hardness results and efficient algorithms for different scenarios.
In particular, we establish both simple and non-trivial reductions
\emph{from} resp.~\emph{to} classic combinatorial problems, and also derive
several new algorithms from scratch 
which may be of interest
beyond the scope of this paper.

In summary, we make the following 
contributions.
For a single waypoint ($k=1$), we show the following:
\begin{enumerate}

\item \textbf{Waypoint routes can be computed efficiently on undirected graphs:}
We establish a connection to the classic disjoint paths problem,
but show that while the 2-disjoint paths problem is notoriously hard
and continues to puzzle researchers~\cite{thore-icalp}, a route via
a single waypoint can in fact be computed very efficiently.

\item \textbf{Waypoints which change the flow size are challenging:}
We show that routing through a single waypoint is NP-hard in general if the 
waypoint changes the flow. This can be seen as an interesting new insight 
into the classic 2-disjoint paths problem as well. 

\item \textbf{Directed links make it hard as well:} While we describe fast algorithms
for undirected networks, the waypoint routing problem turns out to be NP-hard
already for a single waypoint on directed graphs. 

\item \textbf{Supporting absolute distance and stretch constraints
is difficult:} We point out another frontier for the computational
tractability of computing routes through a single waypoint:
the problem also becomes NP-hard if in addition to minimizing the total length of the
route, there are hard distance (or stretch constraints) between the source resp.~destination and the waypoint.
\end{enumerate}

For multiple waypoints (arbitrary~$k$), we show: % the following: % results:
\begin{enumerate}
\item \textbf{Routes through a fixed number of waypoints 
can be computed in polynomial time:} 
This result follows by a reduction to a classic result by 
Robertson and Seymour~\cite{DBLP:journals/jct/RobertsonS95b}.

\item \textbf{Already the decision problem is hard in general:} 
For general~$k$, even on undirected graphs, the decision 
problem (whether a feasible route \emph{exists}) is NP-hard.

\end{enumerate}

Motivated by these results and the fact that the topologies of real-world networks 
(e.g., datacenter, enterprise, carrier networks) are often not arbitrary but feature additional structure, 
we take a closer look at special networks.
\begin{enumerate}
\item \textbf{In reality, there is hope:} We present several algorithms 
to compute shortest waypoint routes on specific graph
families, and in particular, on networks of bounded treewidth.

\item \textbf{An accurate characterization of tractability:} 
We show that it is difficult to go significantly beyond the graph
families studied above, by deriving NP-hardness results on slightly
more general graph families already (graphs of treewidth
three).
\end{enumerate}

An overview of our complexity results 
derived in this paper can be found in Table~\ref{tbl:small}.
We further note that in the following figures, we will draw~$(s,w)$ paths in solid red and~$(w,t)$ paths in solid blue, depicting alternative paths in a dotted style.

\subsection{Paper Organization}

We start in Sec.~\ref{sec:one-waypoint} by considering routing problems via a single waypoint, before studying multiple waypoints in Sec.~\ref{sec:k-waypoints}. 
We then discuss our case study in Sec.~\ref{sec:casestudy},
covering further related work in Sec.~\ref{sec:relwork},
before concluding in Sec.~\ref{sec:conclusion}. 

\begin{comment}

\klaus{story}
\begin{itemize}
	\item General graphs
	\begin{itemize}
		\item 1WP, \Cref{sec:one-waypoint} (undirected P, undirected BW NPC, directed NPC)
		\item~$O(n)$ WPs \Cref{sec:k-waypoints}: undirected (Eulerian) NPC (also Eulerian directed NPC)
		\item constant WPs: \Cref{subsec:constant-undirected}: undirected feasible P
	\end{itemize}
	\item Case Study, sparse Graphs \Cref{sec:casestudy}
	\begin{itemize}
		\item und.~outerplanar feasible, TW3 NPC \Cref{sec:case-tw}
		\item but for BW, cycle is NPC, for both undirected and directed \Cref{sec:cycle}
		\item at least for trees \& BW, P (undirected and directed, if underlying graph is a tree) \Cref{sec:trees}
		\item for directed graphs with BW: also P for DAGs \Cref{sec:dags}
	\end{itemize}
	\item Apply toolbox of transformations \Cref{sec:toolbox}
	\item Restrict first path etc \Cref{sec:restriction}
\end{itemize}

\end{comment}

\section{Routing Via A Waypoint}\label{sec:one-waypoint}

We start by considering the fundamental problem of how to route a flow from~$s$ to
$t$ via a \emph{single} waypoint~$w$.

\subsection{Undirected Graphs Are Tractable}

Many graph theoretical problems revolve around
undirected graphs, and
we therefore also consider them first.
In undirected graphs,
flows can consume bandwidth
capacity in both directions: e.g., a link of capacity
two can accommodate two unit-size flows traversing it both in
opposite directions as well as in the same direction.

Before delving into the details of our algorithms and
hardness results,  
we make some general observations. 
First, we observe that 
a (shortest) route~$(s,w,t)$ can be decomposed into
two segments 
$(s,w)$ and~$(w,t)$. 
While~$(s,w,t)$ can contain loops, the 
two segments 
$(s,w)$ and~$(w,t)$
are \emph{simple paths}, 
without loss of generality:
any loop on a route segment can simply be shortcut.
More generally, we make the following observation.
\begin{observation}
A shortest route (describing a walk) through~$k$ waypoints can be decomposed into 
$k+1$ simple paths~$P_i$ between terminals and waypoints:
$\route=(P_1,\ldots,P_{k+1})$.
\end{observation}

Second, we observe that we can transform
the capacitated problem variant to an uncapacitated
one, by replacing capacitated links with a (rounded-down) number
of parallel, uncapacitated links. Computing
a capacity-respecting walk on the capacitated graph
is then equivalent to computing a link-disjoint
path on the uncapacitated network.\footnote{If parallel links are undesired, each link could
additionally be subdivided by placing an additional
node in the middle, and splitting the link cost in two halfs
accordingly. As discussed later in more details, note however that the resulting topology
may have different properties than the original one (due to the parallel
resp.~subdivided links).}
 
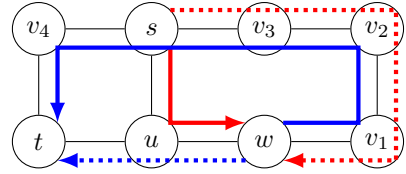
\begin{figure}[!h]
	\centering
	\begin{tikzpicture}[auto]
	\node [markovstate] (s) at (1.5, 1.5) {$s$};
	\node [markovstate] (u) at (1.5, 0) {$u$};
	\node [markovstate] (w) at (3, 0) {$w$};
  \node [markovstate] (t) at (0, 0) {$t$};
  \node [markovstate] (v3) at (3, 1.5) {$v_3$};
	\node [markovstate] (v2) at (4.5, 1.5) {$v_2$};
	\node [markovstate] (v1) at (4.5, 0) {$v_1$};
  \node [markovstate] (v4) at (0, 1.5) {$v_4$};
	%\node [markovstate] (v5) at (-1.5, 0) {$v_5$};
  %\node [markovstate] (v6) at (-1.5, -1.5) {$v_6$};
	%\node [markovstate] (v7) at (0, -1.5) {$v_7$};

	\draw (s) to (u);
	\draw (u) to (w);
  \draw (u) to (t);
	\draw (v3) to (v2);
	\draw (v2) to (v1);
	\draw (v1) to (w);
	%\draw (s) to (v2);
	\draw (v3) to (s);
	\draw (s) to (v4);
	%\draw (s) to (v4);
	\draw (v4) to (t);
	%\draw (v5) to (v6);
	%\draw (v6) to (v7);
	%\draw (v7) to (t);
	
	%\draw [ultra thick, draw=red!200] plot [smooth, tension=0.4] coordinates {(1.6,1.2) (1.65,0.3) (2.75,0.25)};
	\draw [markovedge, ultra thick, draw=red!200] plot coordinates {(1.75,1.25) (1.75,0.25) (2.75,0.25)};
	%\draw [ultra thick, draw=blue!200] plot [smooth, tension=0.4] coordinates { (3.25,0.25) (4.3,0.3) (4.25, 1.25) (1.75,1.35) (0.2, 1.25) (0.1,0.35)};
	\draw [markovedge, ultra thick, draw=blue!200] plot coordinates { (3.25,0.25) (4.25,0.25) (4.25, 1.25) (1.75,1.25) (0.25, 1.25) (0.25,0.25)};
	%\draw [ultra thick, dotted, draw=red!200] plot [smooth, tension=0.4] coordinates {(1.75,1.75) (4.75, 1.65) (4.75, -0.15) (3.25, -.25)};
	\draw [markovedge, ultra thick, dotted, draw=red!200] plot coordinates {(1.75,1.75) (4.75, 1.75) (4.75, -0.25) (3.25, -.25)};
	%\draw [ultra thick, dotted, draw=blue!200] plot [smooth, tension=0.4] coordinates {(2.75,-0.25) (0.25, -0.25)};
	\draw [markovedge, ultra thick, dotted, draw=blue!200] plot  coordinates {(2.75,-0.25) (0.25, -0.25)};

	\end{tikzpicture}
	\caption{In undirected graphs, path segments need to be jointly optimized: greedily selecting a shortest path from~$s$ to~$w$ can force a very long path from~$w$ to~$t$.
Once the solid red~$(s,w)$ path has been inserted first as a shortest path, there is only one option for the solid blue~$(w,s)$ path, resulting in a walk length of~$2+6=8$. A joint optimization leads to the dotted red~$(s,w)$ path and the dotted blue~$(w,t)$ path, with a total length of~$4+2=6$. %\klaus{unify drawing style}
} 
	\label{fig:blocking-und}
\end{figure}

These observations provide us with a 
first idea to compute a shortest route~$(s,w,t)$:
we could simply compute the two optimal paths 
$(s,w)$ and~$(w,t)$ independently.
That is, we could route the first segment from~$s$ to~$w$ along the shortest
path, subtract the consumed bandwidth along the path,
and then compute a shortest feasible path from~$w$ to~$t$ on the remaining graph.
The example depicted in Fig.~\ref{fig:blocking-und} 
shows how this strategy fails. Therefore, we conclude that in an undirected
setting, we need to \emph{jointly} optimize the two paths.

However, the above observations also allow us to compute
an optimal solution: the computation of shortest link-disjoint paths~$(s_1,t_1)$ and~$(s_2,t_2)$
is a well-known combinatorial problem, to which we can directly
reduce the waypoint routing problem by setting~$s_1=s, t_1=s_2=w, t_2=t$.
Unfortunately however, while a recent breakthrough result~\cite{thore-icalp} 
has shown how to compute shortest two disjoint paths
in randomized polynomial time, the result is a theoretical one: 
the order of the runtime polynomial is far from practical. 

\begin{figure}[h]
	\centering
	\begin{tikzpicture}[auto]
	\node [markovstate] (S) at (0, 0) {$S^+$};
	\node [markovstate] (s) at (1.5, 0.5) {$s$};
	\node [markovstate] (t) at (1.5, -0.5) {$t$};
	\node (dots1) at (3,0) {\huge~$G$};
	\node (dots2) at (3,1) {$\dots$};
	\node (dots2) at (3,-1) {$\dots$};
	\node [markovstate] (w) at (4.5, 0) {$w$};
	\node [markovstate] (T) at (6.25,0) {$T^+$};
	
	\draw [markovedge, ultra thick, draw=red!200] (S) to (s);
	\draw [markovedge, ultra thick, draw=blue!200] (S) to (t);
	\draw [dotted, ultra thick] (3,0) ellipse (2.5cm and 1.3cm);
	
	\draw [markovedge, ultra thick, draw=red!200] plot [smooth, tension=0.4] coordinates {(4.75,0.25) (6,0.25)};
	\draw [markovedge, ultra thick, draw=blue!200] plot [smooth, tension=0.4] coordinates {(4.75,-0.25) (6,- 0.25)};

	\end{tikzpicture}
	\vspace{-0.2cm}
	\caption{By adding a super-source~$S^+$, we can reduce the 
	 waypoint routing problem on undirected graphs to a min cost flow problem:
	  As a~$w-t$ flow is also a~$t-w$ flow, we can check if there is a flow of
	   size 2 from~$S^+$ to~$w$. An analogous idea can be used to  reduce the waypoint routing problem to finding two link-disjoint paths 
	 from~$S^+$ to~$T^+$, later reversing the blue path direction in the undirected 
	 case.}

  \label{fig:easy-undirected-1wp}
	\vspace{-0.1cm}
\end{figure}
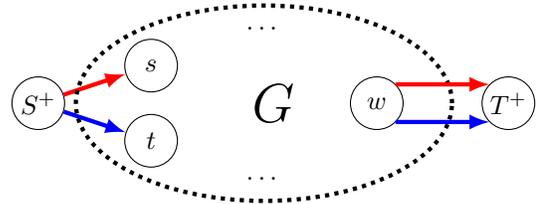

Yet, there is hope: our problem is strictly simpler, 
as the two paths have a common endpoint~$t_1=s_2=w$.
Indeed, the common endpoint~$w$ can be leveraged to employ a 
reduction to an integer flow formulation:
introduce a super-source~$S^+$ and a super-destination~$T^+$, 
connect~$S^+$ to~$s$ and~$t$, and~$T^+$ to~$w$ with two links, 
all of unit capacity, see Fig.~\ref{fig:easy-undirected-1wp}. 

Next, 
%assign all links a unit cost \klaus{change this if we use weights}, and 
solve the minimum cost integer flow problem from~$S^+$ to~$T^+$ with a demand of 2.
By performing flow decomposition and removing~$S^+, T^+$, we obtain an~$s-w$ and a~$w-t$ flow, whose combined length is minimum. Note that in undirected graphs, any~$s-t$ flow can also be interpreted as a~$t-s$ flow.
It is well-known that this flow problem can be solved fairly efficiently: 
for a single source and a single destination, the minimum cost integer flow can be 
solved in polynomial time $O((m\log m) (m+n\log n))$, cf.~\cite[p.~227]{korte2012combinatorial}\footnote{Undirected instances can be turned into directed ones, by replacing each link by two antiparallel directed links. After obtaining a directed solution, flows on antiparallel links can be canceled out, obtaining an undirected solution.}. 

But there exist even better solutions. We can leverage a reduction to a problem
concerned with the computation of 
two (shortest) disjoint paths 
\emph{between the same endpoints~$s$ and~$t$}. 
For this problem, there exists a well-known and
fast algorithm by Suurballe for node-disjoint paths~\cite{DBLP:journals/networks/Suurballe74}: 
it first uses Dijkstra's algorithm to find a first path, 
modifies the graph links, and then runs Dijkstra's algorithm a second time.
It was extended 10 years later to link-disjoint paths by Suurballe and Tarjan~\cite{Suurballe1984}:

\begin{theorem}\label{thm:easy-undirected-1wp}
On undirected graphs with non-negative link weights, the shortest waypoint routing problem can be solved 
 for a single waypoint in time~$O(m \log_{(1+m/n)}n)$.
\end{theorem}

\begin{IEEEproof}
We will make use of Suurballe's algorithm extended to the link-disjoint 
case~\cite{Suurballe1984} in our proof, which solves the following problem in time~$O(m \log_{(1+m/n)} n)$: Given a directed graph~$G=(V,E)$, find two link-disjoint paths from~$s$ to~$t$, 
with~$s,t \in V$, where their combined length is minimum. 
To apply it to the undirected case, we can make use of a standard reduction from undirected 
graphs to directed graphs for link-disjoint paths, replacing every undirected link with five directed links~\cite{DBLP:conf/bonnco/NavesS08}, see Figure~\ref{fig:undtodir}.

\begin{figure}[t]
	\centering
	\begin{tikzpicture}[auto]
	\node [markovstate] (u) at (0, 0) {$u$};
	\node [markovstate] (v) at (2, 0) {$v$};
	
	\node [markovstate] (u1) at (4, 0) {$u$};
	\node [markovstate] (v1) at (8, 0) {$v$};
	\node [markovstate] (uv1) at (6, 0.75) {$x$};
	\node [markovstate] (uv2) at (6, -0.75) {$y$};

	\draw (u) to (v);
	\draw [markovedge] (u1) to (uv1);
	\draw [markovedge] (v1) to (uv1);
	\draw [markovedge] (uv1) to (uv2);
	\draw [markovedge] (uv2) to (v1);
	\draw [markovedge] (uv2) to (u1);

	\end{tikzpicture}
	\caption{By replacing every undirected link with the construction to the right, we can apply algorithms for directed graphs to undirected graphs. Observe that in both cases, the integer flow possibilities between $u$ and $v$ are identical. Regarding link weights, we set the weight from $x$ to $y$ to being the one of the undirected case, with all other four weights being zero.}
	\label{fig:undtodir}
\end{figure}
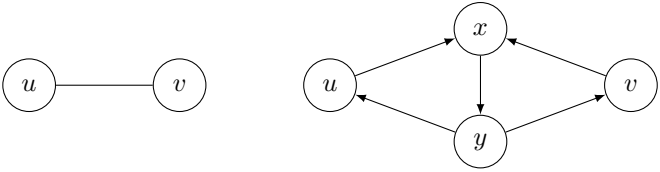

As the flow orientation is not relevant on undirected graphs, we obtain a solution for finding two link-disjoint paths from~$s$ to~$t$ on undirected graphs.

Note that the above applies to unit link capacities, which 
we extend to larger link capacities as follows:
We can apply a standard reduction technique, creating two 
parallel undirected links if the capacity suffices. Observe that 
more than two parallel links do not change the feasibility.

Now add a super-source~$S^+$ and a super-destination~$T^+$ to 
the transformed directed graph, connecting~$S^+$ to~$s$ and~$t$ with links of unit capacity, 
and~$T^+$ to~$w$ with two links of unit capacity, see Fig.~\ref{fig:easy-undirected-1wp}.
 To remove the parallel link property from the graph, nodes are placed on all links, 
 splitting them into a path of length two, scaling path lengths by a factor of two.
In total, the number of nodes and links are still in~$O(n)$ and~$O(m)$, respectively, allowing us to run Suurballe's extended algorithm in~$O(m \log_{(1+m/n)}n)$.
Lastly, by removing~$S^+, T^+$, translating the graph back to be undirected, and scaling the path lengths back, we obtain an~$s-w$ and a~$w-t$ path, whose combined length is minimum. If no solution exist, Suurballe's extended algorithm will notice it during its execution.
\end{IEEEproof}

Thus, we conclude that finding a shortest~$(s,w,t)$ walk is significantly
simpler than shortest two paths~$(s_1,t_1)$,~$(s_2,t_2)$.

\begin{figure}[b]
	\centering
	\begin{tikzpicture}[auto]
	\node [markovstate] (S) at (0, 0) {$S^+$};
	\node [markovstate] (s) at (1.5, 0.75) {$s_1$};
	\node [markovstate] (t) at (1.5, -0.75) {$s_2$};
	%\node (dots1) at (3,0) {\huge~$G$};
	%\node (dots2) at (3,1) {$\dots$};
	%\node (dots2) at (3,-1) {$\dots$};
	\node [markovstate] (w1) at (4.5, 0.75) {$t_1$};
	\node [markovstate] (w2) at (4.5, -0.75) {$t_2$};
	\node [markovstate] (T) at (6.25,0) {$T^+$};
	
	\draw [markovedge, ultra thick, draw=red!200] (S) to (s);
	\draw [markovedge, ultra thick, draw=red!200, dotted] plot [smooth, tension=0.4] coordinates {(0.17,0.34) (1.17,0.75)};
	\draw [markovedge, ultra thick, draw=red!200] (s) to (w2);
	\draw [markovedge, ultra thick, draw=red!200, dotted] (s) to (w1);
	\draw [markovedge, ultra thick, draw=red!200, dotted] plot [smooth, tension=0.4] coordinates {(4.85,0.75) (6.2,0.34)};
	\draw [markovedge, ultra thick, draw=red!200] (w2) to (T);
	\draw [markovedge, ultra thick, draw=blue!200] (S) to (t);
	\draw [markovedge, ultra thick, draw=blue!200, dotted] plot [smooth, tension=0.4] coordinates {(0.17,-0.34) (1.17,-0.75)};
	\draw [markovedge, ultra thick, draw=blue!200] (t) to (w1);
	\draw [markovedge, ultra thick, draw=blue!200, dotted] (t) to (w2);
	\draw [markovedge, ultra thick, draw=blue!200, dotted] plot [smooth, tension=0.4] coordinates {(4.85,-0.75) (6.2,-0.34)};
	\draw [markovedge, ultra thick, draw=blue!200] (w1) to (T);
	%\draw [dotted, ultra thick] (3,0) ellipse (2.5cm and 1.6cm);
	
	%\draw [markovedge, ultra thick, draw=red!200] plot [smooth, tension=0.4] coordinates {(4.75,0.25) (6,0.25)};
	%\draw [markovedge, ultra thick, draw=blue!200] plot [smooth, tension=0.4] coordinates {(4.75,-0.25) (6,- 0.25)};
	\end{tikzpicture}
	%\vspace{-0.2cm}
	\caption{Extending the idea of Figure~\ref{fig:easy-undirected-1wp} to two link-disjoint paths can fail, as shown in this figure: Instead of finding a $S^+,s_1,t_1,T^+,t_2,s_2,S^+$ path (depicted in dotted red and blue), the output could be to first visit $t_2$ and then $t_1$ second, never visiting $t_2$ again after (depicted in solid red and blue).}
  \label{fig:easy-undirected-2wp}
\end{figure}
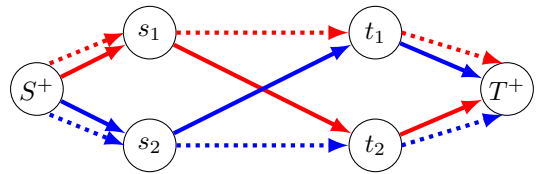

\noindent \textbf{Remark.} One might wonder whether the above approach
can also be employed to efficiently compute 2-disjoint paths $(s_1,t_1)$, $(s_2,t_2)$,
e.g., using a construction similar to the one outlined in Figure~\ref{fig:easy-undirected-2wp}.
%\stefan{Klaus maybe it is better we replace in the figure: $s$ with $s_1$, $w_1$ with $t_1$,
%$t$ with $s_2$, $w_2$ with $t_2$?}
The problem with this idea is that $s_1$ may be matched to $t_2$
and  $s_2$ to $t_1$. Indeed, the problem of finding
two disjoint paths from $\{s_1,s_2\}$ to $\{t_1,t_2\}$ where the matching is subject
to optimization, is significantly simpler (and can be solved, e.g., using a flow algorithm).

\subsection{Flow Size Changes Make it Hard} 

We have assumed so far that traffic rates are not changed at waypoints.
However, there are many scenarios where waypoints increase or
decrease the bandwidth demand. For example, the addition of an 
encapsulation header will increase the packet sizes whereas a
wide-area network optimizer may compress the packets. 
Unfortunately, it turns out that computing routes through 
a single waypoint
which changes the bandwidth is much harder than routing through
waypoints which do not: 

\begin{theorem}\label{thm:und-cap}
On undirected graphs in which waypoints are not flow-conserving, 
computing a route through a single waypoint 
is strongly NP-complete.
\end{theorem}
An NP-complete problem is strongly NP-complete, if the input can be restricted to numbers in unary representation.
\begin{IEEEproof}
Reduction from the strongly NP-complete 2-splittable 
flow problem: 
Given an undirected graph~$G$ with link capacities, are there two paths to route the flow from~$S^+$ to~$T^+$ s.t.~the flow is maximized?
Koch and Spenke showed in~\cite{KOCH2006338} that
determining whether the maximum throughput is 2 or 3 in 
  the 2-splittable 
flow problem is strongly NP-hard on undirected graphs with link capacities of 1 or 2.

Our reduction will be from the corresponding decision problem, i.e., 
does a flow of size~$3$ exist? 
Assume for ease of construction that~$s:=S^+=:t$ and~$w:=T^+$. 
As all link capacities are either 1 or 2, we only need to check the variants~$d_0 \in \left\{1,2\right\}, d_1 \in \left\{1,2\right\}$ of the capacitated waypoint routing problem for feasibility.
Therefore, if there was a polynomial algorithm for the capacitated waypoint routing problem on undirected graphs, we would also obtain a polynomial algorithm for the initial problem. Lastly, the capacitated waypoint routing problem is clearly in NP.
\end{IEEEproof}

\subsection{Directions Are Challenging As Well}
But not only waypoints changing the flow sizes turn the problem hard quickly:
in a \emph{directed} network, already the problem of finding a
\emph{feasible}
waypoint route is NP-hard, even if waypoints are flow conserving.
\begin{theorem}\label{thm:hard-directed}
On directed graphs, the waypoint routing problem is strongly NP-complete for a single waypoint.
\end{theorem}
\begin{IEEEproof}
Our proof is by a reduction from the strongly NP-complete 2-link-disjoint paths problem~\cite{DBLP:journals/tcs/FortuneHW80}:
Given two node pairs~$(s_1,t_1)$,~$(s_2,t_2)$ in a directed graph~$G=(V,E)$, are there two link-disjoint paths~$P_1=s_1,\dots,t_1$,~$P_2=s_2,\dots,t_2$?
\begin{comment}
The NP-completeness of this problem follows from the directed subgraph 
homomorphism problem~\cite{DBLP:journals/tcs/FortuneHW80,DBLP:conf/bonnco/NavesS08}
\end{comment}

We perform a reduction of all problem instances~$I$ of the 2-link-disjoint paths problem in graphs~$G$ to instances~$I'$ in graphs~$G'$ as follows:
 Create a (waypoint) node~$w$, and add the directed links~$(t_1,w)$ and~$(w,s_2)$, see Fig.~\ref{fig:hard-directed}. 
\begin{figure}[bthp]
	\centering
	\begin{tikzpicture}[auto]
	\node [markovstate] (t1) at (0, 0) {$t_1$};
	\node [markovstate] (w) at (2, 1) {$w$};
	\node [markovstate] (s2) at (4, 0) {$s_2$};
	\draw [dotted, ultra thick] (2,0) ellipse (4cm and 0.6cm);
	\node (dots1) at (1.8,0) {\huge~$G$};
	
	\draw [markovedge] (t1) to (w);
	\draw [markovedge] (w) to (s2);

	%\draw [markovedge, ultra thick, draw=red!200] plot [smooth, tension=0.4] coordinates {(-0.25,0.25) (-1.75,0.25)};
	%\draw [markovedge, ultra thick, draw=blue!200] plot [smooth, tension=0.4] coordinates {(-1.75,-0.25) (1.75,-0.25)};
	%
	\end{tikzpicture}
	\vspace{-0.2cm}
	\caption{By adding the waypoint~$w$ on a directed path between~$t_1$ and~$s_2$, every feasible solution of the waypoint routing problem must be a concatenation of two walks~$s_1,\dots,t_1,w$ and~$w,s_2,\dots,t_2$.
	}
	\label{fig:hard-directed}
\end{figure}
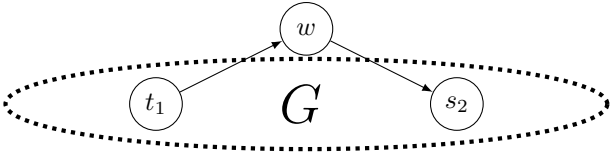

To finish the construction of the waypoint routing problem in~$I'$, set~$s:=s_1$ and~$t:=t_2$: Is there a route from~$s$ via~$w$ to~$t$, using every link only once? %\klaus{capacities?}

If~$I$ is a yes-instance,~$I'$ is a yes-instance as well, by joining the paths~$P_1, P_2$ via the directed links~$(t_1,w)$ and~$(w,s_2)$.
Next, we show that if~$I$ is a no-instance,~$I'$ is a no-instance as well:
First, observe that to traverse~$w$ in~$G'$ starting from~$s$, the only option is 
via traversing both links~$(t_1,w)$ and~$(w,s_2)$, successively in that order. 
Thus, assume for the sake of contradiction that~$I'$ is a yes-instance with a link-disjoint walk~$W=s,\dots,t_1,w,s_2\dots,t$. Then, we can also create two link-disjoint walks~$W_1=s_1,\dots,t$ and~$W_2=s_2,\dots,t_2$ in~$I$ by removing both links~$(t_1,w)$ and~$(w,s_2)$ from~$W$. Removing the loops in~$W_1$ and~$W_2$ results in paths~$P_1$ and~$P_2$ solving~$I$, a contradiction.

To conclude, observe that the reduction can be performed in polynomial time, and as the problem is clearly in NP, the problem is NP-complete.
\end{IEEEproof}

\subsection{Another Complexity: Distance Constraints}

Another problem variant arises if 
we do not only want to find a feasible (or shortest) path from
$s$ via~$w$ to~$t$, but also have \emph{hard constraints
on the distance} or stretch from~$s$ to the waypoint,
or from the waypoint to the destination. 

\begin{theorem}\label{thm:hard-hard-undirected}
%The problem of 
Finding a feasible path from~$s$ to~$t$
via~$w$ subject to distance constraints between two consecutive
%nodes
nodes from $s,w,t$ is strongly NP-complete
%already on undirected graphs. 
on undirected graphs.
\end{theorem}
\begin{IEEEproof}
This follows by reduction due to the 
hardness of finding 2 link-disjoint paths under
a \emph{min max objective}.
Li et al.~\cite{li1990complexity9}
showed that given a graph~$G = (V, E)$ and two nodes~$s'$ and~$t'$, 
the problem of finding two
disjoint paths from~$s'$ to~$t'$ such that the length of the longer path is minimized
is strongly NP-complete, even with unit link weights. This implies that the waypoint routing
problem is strongly NP-complete as well, by setting~$s=t=s'$
and~$w=t'$. 
\end{IEEEproof}

Recall that the directed case of a single waypoint was already hard without distance constraints, see \Cref{thm:hard-directed}.

We note that Itai et al.~\cite{DBLP:journals/networks/ItaiPS82} showed the two link-disjoint path problem with distance constaints to be NP-complete on directed acyclic graphs, using exponential link weights (polynomial in binary representation) in their construction.
However, as we will show later, the distance constrained directed waypoint routing problem is polynomially time solvable on DAGs, even for arbitrarily many waypoints.

\section{Routing Via Multiple Waypoints}\label{sec:k-waypoints}

The advent of more complex network services requires the routing of traffic through \emph{sequences of (multiple) waypoints}. 
Interestingly, and despite the 
numerous hardness results derived for a single waypoint in the previous
section, we will see that it is still possible to derive some polynomial-time
algorithms even for multiple waypoints. 

\subsection{Possible For a Fixed Number of Waypoints...}\label{subsec:constant-undirected}

Interestingly, the~$k$-waypoint routing problem is tractable 
when the number of waypoints is constant:

\begin{theorem}\label{thm:easy-undirected-kwp}
On undirected graphs, one can decide in polynomial time $O(m^2)$ whether a feasible route
through a fixed number of waypoints exists.
\end{theorem}

\begin{IEEEproof}
The proof follows by application of~\cite{DBLP:journals/jct/KawarabayashiKR12}, building upon the seminal work of Robertson and Seymour~\cite{DBLP:journals/jct/RobertsonS95b}:
the authors show that for any
fixed~$k$, the
$k$-link-disjoint path problem can be decided in polynomial-runtime
of~$O(n^2)$ on undirected graphs. We can apply their result by asking for link-disjoint paths connecting the waypoints in successive order.
It only remains to set all link capacities to one: 
To do so, we divide the links into parallel links, their number bounded by $k \in O(1)$, even if the capacity is higher. 
Then, we place a node on every link, obtaining $O(n+km) \in O(m)$ nodes.
\end{IEEEproof}
An analogous result holds for bidirected graphs~\cite{Foerster2017}.

\begin{table*}[t!]

\centering
\setlength\extrarowheight{4pt}
\begin{threeparttable}
%\resizebox{\linewidth}{!}{%
\begin{tabular}{>{\centering\arraybackslash}p{0.8in}>{\centering\arraybackslash}p{1.2in}>{\centering\arraybackslash}p{1.3in}>{\centering\arraybackslash}p{1.8in}>{\centering\arraybackslash}p{1.35in}}
\multicolumn{1}{c|}{\textbf{$\#$ Waypoints}}
&
  \textbf{Feasible Algorithms}
&
  \multicolumn{1}{c}{\textbf{Feasible Hardness}}
&
  \textbf{Demand Change Optimal Algorithms}
&
  \textbf{Feasible Hardness}
\\ \hline\hline

\multicolumn{1}{c|}{\textbf{Arbitrary}} & \textbf{P}: Outerplanar ($\tw \leq2$) \Cref{corr:und-ordered-outerpl} & \textbf{Strongly NPC}: $\tw \leq3$ \Cref{thm:und-ord-arb-tw3} & \textbf{P}: Tree (equivalent to $\tw$ of $1$) Observation~\ref{obs:trees} & \textbf{NPC}: Unicyclic ($\tw \leq 2$) \Cref{thm:und-ord-arb-tw3}
\\ \hline
\multicolumn{1}{c|}{\textbf{Constant}} & \textbf{P}: General graphs \Cref{thm:easy-undirected-kwp} &  \textbf{P}: General graphs \Cref{thm:easy-undirected-kwp} & \textbf{P}: Constant treewidth $\tw \in O(1)$ \Cref{thmconstconst} & \textbf{Strongly NPC}: General graphs \Cref{thm:und-cap}
\\ \hline

%{\textbf{P} (Thm.~\ref{thm:fptund})
\end{tabular}
%}

\end{threeparttable}
\vspace{0.1cm}
\centering \caption{Overview of the Complexity Landscape for Waypoint Routing in Special Undirected Graphs. }
\label{tbl:small1}
\vspace{-0.6cm}
\end{table*}
\subsection{... Hard Already on Eulerian Graphs}
%%%%%%%%%%%%%%%%%%%%%%%%%%%the following proof is wrong%%%%%%%%%%%%%%%%%%%%%%%%%%%%%%%%

While polynomial-time solutions exist for fixed
$k$ on general graphs, we now show that for general~$k$, % unfortunately, 
the problem is computationally intractable already 
%on very 
%simple graphs: % as well:
%restricted graphs, namely 
on undirected Eulerian graphs
(graphs on which routing problems are often simple),
where all nodes have even degree.\footnote{Beyond constant $\nowp$, there is an $O(n^2)$ link-disjoint path  algorithm on Eulerian graphs that allows $\nowp \in O((\log \log \log n)^{\frac{1}{2}-\varepsilon})$, for any $\varepsilon >0$~\cite{DBLP:journals/combinatorica/Kawarabayashi015}. For planar Eulerian graphs, this also extends to $\nowp \in O((\log n)^{\frac{1}{2}-\varepsilon}$).}

\begin{theorem}\label{thm:euler-np-hard}
The waypoint routing problem is strongly NP-complete on undirected Eulerian graphs.
\end{theorem}
\begin{IEEEproof}
We briefly introduce some notions of the problem that we will use for the reduction.
The link-disjoint path problem can also be formulated via a supply graph $G=(V,E)$, which supplies the links to route the paths, and a demand graph $H=(V,E(H))$, whose links imply between which nodes there is a demand for a path.
I.e., $\left\{(s_1,t_1),\dots,(s_k,t_k)\right\}=E(H)$.
The union of both graphs is defined as $(V,E \cup E(H))$.
As this notation is rather uncommon in a networking context, we provide a small introductory example in Figure~\ref{fig:Eulerian-Demand}.

\begin{figure}[h]%
\vspace{-0.5cm}
\begin{center}
	\begin{tikzpicture}[auto]
	\node [markovstate] (v1) at (-2, 0) {$v_1$};
	\node [markovstate] (v2) at (0, 0) {$v_2$};
	\node [markovstate] (v3) at (2, 0) {$v_3$};
	\node [markovstate] (v4) at (4, 0) {$v_4$};
	\node [markovstate] (v5) at (6, 0) {$v_5$};

	\draw (v1) to (v2);
	\draw (v2) to (v3);
	\draw (v3) to (v4);
	\draw (v4) [bend left] to (v2);
	\draw (v4) to (v5);
	\draw (v4) [bend right, dotted, thick] to (v2);
	\draw (v4) [bend right, dotted, out=340, in=200, thick] to (v2);
	\draw (v5) [bend right, dotted, thick] to (v1);
	\draw (v5) [bend right, dotted, out=320, in=220, thick] to (v1);
	\draw (v5) [bend right, dotted, out=340, in=200, thick] to (v1);

	%\draw [markovedge] (uv2) to (u1);

	\end{tikzpicture}
\end{center}
\vspace{-0.5cm}
\caption{Here, the supply graph $G$ consists of  $V=\left\{v_1,v_2,v_3,v_4,v_5\right\}$ and $E=\left\{(v_1,v_2),(v_2,v_3),(v_3,v_4),(v_4,v_5),(v_2,v_4)\right\}$, drawn in solid. The demand graph $H$ has the same node set $V$, but its links $H(E)$ contain two links from $v_2$ to $v_4$ and three links from $v_1$ to $v_5$, drawn dotted. Note that $(V,E \cup E(H)$ is planar and Eulerian.
Still, the link-disjoint path problem given by the supply and demand graph is not solvable in this instance, 
e.g., only one path can be routed from $v_1$ to $v_5$. If the demand graph did not contain any parallel links, 
both paths could be routed in a link-disjoint fashion.
}
\label{fig:Eulerian-Demand}%
\end{figure}
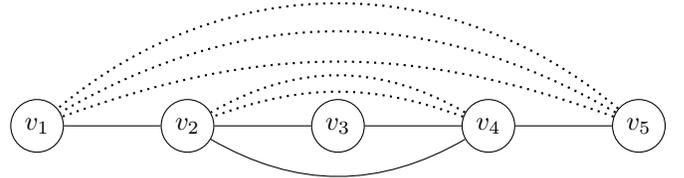

\begin{comment}
\stefan{personally i like it more when we first show hardness and then conclude with the problem being in NP: so we can start with the interesting not the boring part. but that is a minor comment}
As the problem clearly belongs to NP, we need to show its NP-hardness, which we will show via a reduction from the link-disjoint path problem on Eulerian graphs: 
Given a graph~$G=(V,E)$ with~$k$ source-destination pairs~$\left\{(s_1,t_1),\dots,(s_k,t_k)\right\}=E(H)$, are there~$k$ link-disjoint paths~$P_1,\dots,P_k$ in~$G$, from~$s_1$ to~$t_1$,~$\dots$, from~$s_k$ to~$t_k$, respectively?
\stefan{i dont understand this proof. what is H?}

The link-disjoint path problem was shown to be NP-complete in both directed or undirected rectangular grids~$G$, even if the union of~$G$ with the demand graph~$H=(V,E(H))$ is Eulerian~\cite{DBLP:journals/dam/Marx04}. \stefan{i dont understand: what is a demand graph? what is a union?}
\end{comment}

We now reduce from the strongly NP-complete problem of finding link-disjoint paths where the union of the supply and the demand graph is Eulerian~\cite{DBLP:journals/dam/Marx04}.
Our polynomial reduction construction 
%from~$G$ 
of an instance~$I$ 
to an undirected graph~$G'=(V',E')$ proceeds as follows: 
We first initialize~$V'=V$ and~$E'=E \cup E(H)$.
Next, we add a new \emph{center} node~$v$ to~$G'$, containing~$s,t$.
For simplicity, we will assume that $v$ also contains the~$k-1$ waypoints~$w_4,w_8, w_{12},\dots,w_{4(k-1)}$; 
those can also be moved to small cycles connected to $v$.
Next, for~$1 \leq i \leq k$, we define the remaining waypoints as follows:~$w_{4i-3}=s_i,w_{4i-2}=t_i,w_{4i-1}=s_i$. We also add two links between~$v$ and each~$s_i$ to~$E'$,~$1 \leq i \leq k$.
I.e., our waypoint problem is now an instance~$I'$: Start in the center node~$v$, go to~$s_1$, then to~$t_1$, back to~$s_1$, then to~$v$; then proceed similarly for~$s_2$~$\dots$, 
to~$s_k$, and ending at~$v$.
We note that new graph is still Eulerian.
Again, in the same spirit as before, we can split the corresponding demand links, possibly twice, moving waypoints there, preserving the restriction of one waypoint per node and removing all parallel links.

We start with the easier case, showing that if~$I$ is a yes-instance,~$I'$ is as well:
We can take the~$k$ disjoint paths of the solution of~$I$ in~$G$, add the~$k$~$(t_i,s_i)$ paths via the links of~$H(E)$, and lastly connect the waypoint path-segments in order via the incident links of~$v$.

It remains to show that if~$I'$ is a yes-instance,~$I$ is as well.
First, if the~$k$~$(t_i,s_i)$ paths use links outside of~$E(H)$, we can alter the solution s.t.~only the links of~$E(H)$ are used for the~$k$~$(t_i,s_i)$ paths: Assume for some~$i$, that the link~$(t_i,s_i) \in E(H)$ is not used for the path~$P$ from~$t_i$ to~$s_i$, but is rather part of another walk~$W_j$ from some~$w_j$ to~$w_{j+1}$ (if the link is not used at all, the swap can be done directly). 
Then we can swap in $W_j$ the link~$(t_i,s_i) \in E(H)$ with~$P$.
Second,~$v$ has a degree of~$2k$. 
Because~$s,t$ and~$w_{4i-2}=t_i$ are pairwise non-consecutive waypoints, 
and no walk between any~$s_i,t_i$ or~$t_i,s_i$ can use the 
links incident to~$v$, they must be used for all subwalks starting and ending at~$v$. 
Thus, the subwalks between the~$k$~$s_i$ and~$t_i$ (which can be simplified to paths) will now only use links already present in~$G$. Lastly, observing that the problem is in NP finishes the proof.
\end{IEEEproof}
A directed graph is called Eulerian, if for each node $u$ holds: The in- and out-degree of $u$ are identical.
Marx also showed in~\cite{DBLP:journals/dam/Marx04} the (implicitly strong) NP-completeness of the directed case.
Thus, we can apply analogous proof arguments.% and obtain
\begin{corollary}
The ordered waypoint routing problem is strongly NP-complete on directed Eulerian graphs
\end{corollary}
For the case of bidirected graphs, which are a subset of directed Eulerian graphs, optimally solving the ordered waypoint routing problem is NP-complete~\cite{Foerster2017}, but the hardness of the feasibility variant is an open question.

\section{Exploring Computational Tractability in Special Networks}\label{sec:casestudy}

Computer networks often have very specific structures: for example,
many data centers are highly structured (e.g., based on Clos topologies~\cite{clos}), but also enterprise and router-level AS topologies for example, while being less symmetric, often 
come with specific properties (e.g., are sparse). 
In this light, the results derived so far may be too conservative:
in practice, much faster algorithms may be possible which are tailored
toward and leverage the specific network structure. 
Accordingly, in this section we explore the waypoint routing
problem on specific graph families. In particular, we are interested
in sparse graphs. We conducted a small empirical study using 
Rocketfuel topologies~\cite{rocketfuel}
and Internet Topology Zoo graphs~\cite{topologyZoo}, and found that they
often have a low path diversity:
almost half of these graphs 
are \emph{outerplanar}, and one third are \emph{cactus graphs}:
\begin{itemize}
	\item a graph is outerplanar if it has a planar drawing s.t.~all vertices are on the outer face of the drawing~\cite{Chartrand1967}
	\item a graph is a cactus graph if any two simple cycles share at most one node~\cite{geller1969reconstruction} (every cactus graph is outerplanar)
\end{itemize}

\subsection{Exact Polynomial-Time Algorithms}

\noindent \textbf{Tree networks.}
On tree networks, paths between two given nodes are unique,
and finding shortest walks hence trivial: simply compute a shortest
path for each path segment (recall: a simple path), one-by-one.
If this walk is feasible, it is optimal; if not, no solution exists.
Note that this also holds if waypoints change the flow rate, and for directed graphs, when the underlying undirected graph is a tree.

\begin{observation}\label{obs:trees}
The shortest waypoint routing problem with demand changes can be solved in polynomial time on trees and DAGs.
\end{observation}

\noindent \textbf{DAGs.} A similar results still holds on Directed Acyclic Graphs (DAGs).
When making a choice for the path to the next waypoint, we can use a simple greedy algorithm: Any link that we use will never be used for a later path (due to the acyclic property).
Hence, we can also minimize distance constraints for trees and DAGs.

In comparison, the link-disjoint path problem is polynomialy solvable for a fixed number of link-disjoint paths on DAGs~\cite{DBLP:journals/tcs/FortuneHW80}, but NP-complete in general already on planar DAGs~\cite{DBLP:journals/dam/Vygen95}.

\begin{observation}\label{obs:differ}
There are graph families for which the waypoint routing problem can be solved efficiently while the disjoint paths problem cannot.
\end{observation}
%As thus, in g
Regarding parametrized complexity: While an $n^{O(\nowp)}$ algorithm exists for DAGSs, the link-disjoint path problem on DAGs is W$\left[1\right]$ hard~\cite{DBLP:journals/siamdm/Slivkins10}, i.e., unlikely fixed-parameter tractable in $\nowp$.

\noindent \textbf{General Observations and Reductions.}
On the other hand, leveraging our connection to disjoint path problems
again, we can also make the following observation:
\begin{observation}
For any graph family on which the~$k+1$ disjoint paths problem
is polynomial-time solvable, we can also find a route through
$k$ waypoints in polynomial time on graphs of unit link capacity.
\end{observation}

Thus, it immediately follows from~\cite{DBLP:journals/jct/Bang-Jensen91}
that the single waypoint routing problem is polynomial time solvable on semicomplete directed graphs, where a directed graph is called semicomplete, if there is at least one directed link between every pair of nodes. 

Another case are directed graphs with constant independence number $\alpha$, where $\alpha = \alpha(G)$ denotes the maximum size of an independent set in $G$. Then, for constant $\alpha, \nowp \in O(1)$, a polynomial time algorithm exists~\cite{DBLP:journals/jct/FradkinS15}.

Having a well-connected graph helps as well: On random undirected graphs $G$, where the set of $2\nowp$ endpoints are chosen by an adversary (e.g., to compute a waypoint routing), it holds with high probability that the $\nowp$ paths exists, if $\nowp \in O(n/\log n)$ and the minimum degree of $G$ is some sufficiently large constant.
The paths can be constructed in randomized time of $O(n^3)$~\cite{DBLP:journals/cpc/FriezeZ00}.
Similar results also hold on Expander graphs~\cite{DBLP:journals/siamcomp/Frieze00}.

\noindent\textbf{Bounded Treewidth Graphs I.} For a further example, on bounded treewidth graphs,
and as long as the number of waypoints $k$ is logarithmically 
bounded, 
the problem is polynomial time solvable, 
because the link-disjoint paths problem is polynomial time solvable:
For a treewidth decomposition of width $\leq \tw$ and $k$ link-disjoint paths, Zhou et al.~\cite{DBLP:journals/algorithmica/ZhouTN00} provide an algorithm with a runtime of 
\vspace{-0.1cm}
\begin{equation}
O\left(n((k+\tw^2)k^{\tw(\tw+1)/2}+k(\tw+4)^{2(\tw+4)k+3}\right)~.
\label{eq:zhou}
\vspace{-0.1cm}
\end{equation}
%\vspace{-0.2cm}
As a constant-factor approximation of treewidth decompositions can be obtained in polynomial time~\cite{6686186}, also beyond constant treewidth, it is therefore possible to solve the waypoint routing problem for any values of $t$ and $k$ s.t.~\Cref{eq:zhou} stays polynomial.
E.g., $\tw,k \in O(\sqrt{\log n/ \log \log n})$, due to
$f(n)^{g(n)}=\exp(\ln(f(n)^{g(n)}))=\exp(g(n)\ln(f(n)))$. This idea can also be extended to polylogarithmic functions $f(n),g(n) \in \text{polylog}(n)$, obtaining quasi-polynomial runtimes of $2^{\text{polylog}(n)} \in\text{QP}$.
Quasi-polynomial algorithms fit sort of in between polynomial and exponential algorithms and it is widely believed that NP-complete problems are not in QP~\cite{DBLP:conf/aussois/Woeginger01}.

Unit capacities can be modeled by introducing parallel links 
and in particular subdividing them by placing auxiliary nodes in the center,
increasing the $\tw$ only by a constant factor.

We thus obtain the following corollary, which does not find shortest routes and is not applicable to demand changes:
\begin{corollary}
In undirected graphs with a treewidth of $ \tw$ and $\nowp$ waypoints, we can solve the waypoint routing problem in polynomial time for the following combinations:
\begin{itemize}
	\item Constant $\tw \in O(1)$, logarithmic $\nowp \in O(\log n)$
	\item $\tw \in O(\sqrt{\log n})$, constant $k \in O(1)$
	\item $\tw, \nowp \in O\left(\sqrt{\log n/ \log \log n}\right)~.$
\end{itemize}
In quasi-polynomial time, we can solve:
\begin{itemize}
	\item $\tw, \nowp \in \text{polylog}(n)~.$
\end{itemize}
\end{corollary}
Nonetheless, note that the non-parallel unit capacity observation is of limited use in general:
for a negative example, an outerplanar graph requires nodes to touch the outer face, 
however, this property will be lost during the graph transformation. 
Yet, as we will show in the following, solutions for outerplanar graph still exist, even in 
arbitrarily capacitated networks.
We note that outerplanar graphs have a treewidth of $\tw \leq 2$.

\noindent \textbf{Outerplanar Graphs.}
We first prove the following lemma.
%helper lemma.

\begin{lemma}
Let~$\mathcal{I}$ be 
a class of \WRP s.t.
\begin{enumerate}
\item the graph~$G$ is planar (w.l.o.g.~we have a planar drawing),
\item the maximum capacity is $c_{\max}$, w.l.o.g.~$n \geq c_{\max} \in \mathbb{N}$,
\item $s,t$ and all waypoints touch the outer face~$\mathcal{F}$ of~$G$,
\item for every node~$v\not\in\mathcal{F}$, 
$\Sigma_{e\colon\{u,v\}\in E(G)}c(e)$ is even.
\end{enumerate}
Then the \emph{feasibility} of
the ordered waypoint routing problem in the class~$\mathcal{I}$ is %polynomial time decidable.
decidable in time $O(n^2)$, with the explicit construction taking $O(m^2 \cdot c_{\max}^2)$. 
\end{lemma}
\begin{IEEEproof}
Let~$I\in \mathcal{I}$ be an instance of the problem. 
Suppose~$s,t$ are the source and terminal and
$w_1,\ldots,w_k$ are waypoints. Define~$w_0=s,w_{k+1}=t$.  We
construct an equivalent instance of the link-disjoint paths problem as
follows.
Replace each link~$e=\{u,v\}$ with capacity~$c$ by~$c \leq c_{\max}$ links with capacity~$1$,
then subdivide those links once, i.e., the number of nodes is in $O(m \cdot c_{\max})$. 
In the newly created instance of
link-disjoint paths problem:
\begin{enumerate}
\item The input graph is planar,
\item all terminal pairs touch the outerface,
\item degree of every node, not in the outerface, is even.
\end{enumerate} 
If only condition 1) and 2) hold, the problem is NP-hard~\cite{DBLP:journals/combinatorica/Schwarzler09}.
But for this class of link-disjoint paths problem, there
are polynomial time  algorithms~\cite{Becker1986} with the following properties: Let $b$ be the number of nodes on the outer face and $n'$ be the total number of nodes. The feasibility of the link-disjoint path problem can be tested in $O(bn')$ and constructing the paths can be done in $O(n'^2)$
which gives us the desired
polynomial time solutions for the original problem.
\end{IEEEproof}

This directly implies the following result.
\begin{corollary}\label{corr:und-ordered-outerpl}
In outerplanar graphs with a maximum link capacity of $c_{\max}$, the 
waypoint routing problem is decidable in time $O(n^2)$, with an explicit construction obtainable in
time~$O\left(m^2 \cdot \min \left\{n^2,c_{\max}^2 \right\}\right)$.
\end{corollary}
A solution to the shortest waypoint routing problem cannot obtained via the same reduction: 
Brandes et al.~\cite{Brandes1996Edge--5975} showed the minimum total length link-disjoint path problem to be NP-hard on graphs satisfying the three conditions mentioned above, already when the maximum degree is at most 4.

For bidirected cactus graphs of constant capacity, the ordered waypoint routing problem can be optimally solved in polynomial time~\cite{Foerster2017}.

\begin{comment}
As the algorithm by provided by Becker and Mehlhorn~\cite{Becker1986} is constructive, we thus also obtain the routing paths.
\end{comment}

\noindent \textbf{Bounded Treewidth Graphs II.}
Let us quickly recap the results on bounded treewidth $\tw$ found so far:
\begin{enumerate}
	\item For constant $\tw$, we can compute walks for $\nowp \in O(\log n)$ waypoints, but those walks will not be optimal (shortest) and the flow has to be of unit size. The same holds for outerplanar graphs (a class with $\tw=2$) for $\nowp \in O(n)$.
	\item For $\tw=1$ ($\equiv$ trees), we can compute shortest walks with demand changes, even for $\nowp \in O(n)$.
\end{enumerate}

As pointed out in the beginning of this section, many network topologies have low treewidth, , especially in the wide-area and enterprise context (e.g., the Rocketfuel and Topology Zoo networks~\cite{rocketfuel}).
We now tackle a problem we thus deem to be realistic:
in practice, the number of waypoints visited by a given flow is likely to be a small constant.

\begin{theorem}\label{thmconstconst}
In undirected graphs with bounded treewidth $\tw \in O(1)$ and a fixed number $\nowp \in O(1)$ of waypoints, we can solve the shortest waypoint routing problem with demand changes 
 in a  runtime of $O(n)$. 
\end{theorem}
\begin{IEEEproof}
Our proof will be via dynamic programming of a nice tree decomposition~\cite{bodlaender1988dynamic} $\mathcal T =(T,X)$ of $G$ as follows:
Using the ideas and terminology of Kloks~\cite{DBLP:books/sp/Kloks94}, each bag of $\mathcal T$ is either a \emph{leaf bag}, a \emph{forget bag} (one node is removed from the separator), an \emph{introduce bag} (a node is added), or a \emph{join bag} (its two children $q_1, q_2$ contain the same nodes).

For bags $\bagname$, we thus define signatures $\sigma_\bagname$, representing already computed solutions of $\bagname$, such that by dynamically programming $\mathcal T$ bottom-up, we obtain an optimal walk $\mathcal W$ at the root bag of $\mathcal T$, if such a $\mathcal W$ exists.

In every optimal solution $\mathcal W$, each path from a $w_i$ to a $w_{i+1}$ will cross each separator $\bagname$ of $G$ at most $\tw$ times. 
{Due to optimality, these individual paths will traverse every node at most once.}
Hence, a signature $\sigma_b$ only needs to represent the at most $\nowp \cdot \tw$ crossings (endpoints) of partial paths through the subgraph of $\bagname$, and the link utilizations these paths use in $E(\bagname)$.
{We additionally store if a path, for from $w_i$ to $w_{i+1}$, with only one endpoint in the signature, contains either $w_i$ or $w_{i+1}$. Note that at most one such path each will exist at any time due to optimality.}
Due to $\nowp, \tw \in O(1)$, we have only $O(1)$ different possible signatures for each bag $\bagname$, with each signature containing only $O(1)$ elements.
{As common, we assume that we can perform standard operations (additions, comparisons etc.) of numerical values in constant time, else, an extra logarithmic factor needs to be included in the total runtime.}
We now present the required algorithms for the induction. 
\begin{itemize}
	\item \textbf{Leaf bags $\bagname$:} In constant time, we can generate all valid signatures, containing at most $\nowp$ paths (each without any links). The only restriction is that if $v \in V(\bagname)$ is a waypoint $w_i$, its paths to $w_{i-1}$ and $w_{i+1}$ must exist.
	\item \textbf{Forget bags $\bagname$:} Let $v$ be the node s.t.~for the child $q$ of $\bagname$ holds: $V(q) \setminus \left\{v\right\}=V(\bagname)$. If $v$ is not a waypoint, then the valid signatures of $\bagname$ are exactly those of $q$ which do not use $v$ as endpoints. If $v$ is a waypoint $w_i$, then additionally must hold: $v$ must be an endpoint of a path from $w_{i-1}$ and the endpoint of a path to $w_{i+1}$.
	\item \textbf{Join bags $\bagname$:} We first $1)$ describe the program and then $2)$ prove its correctness. $1)$: Given two valid signatures of $\bagname$'s children $q_1,q_2$, we perform all possible concatenations, of endpoints of paths for the same $w_i$ to $w_{i+1}$, at the separator nodes $V(\bagname)$, checking $a)$ that the union of the link utilizations in $E(\bagname)$ respect the link capacities and {$b$) that no loops are created (we know the endpoints of each (sub-)path and the their link utilizations in $E(b)$, if they share a link outside $E(b)$, a signature of minimum size will not)}, which results in valid signatures $\sigma_\bagname$ of $\bagname$. $2)$: Assume we missed some valid signature  $\sigma_\bagname$ of $\bagname$: Given $\sigma_b$, we split the paths across the separator, resulting in valid signatures $\sigma_{q_1}, \sigma_{q_2}$ and their subpaths, a contradiction. %\klaus{picture?}
	For an illustration of this procedure, we refer to Figure~\ref{fig:Join-Split}.
	
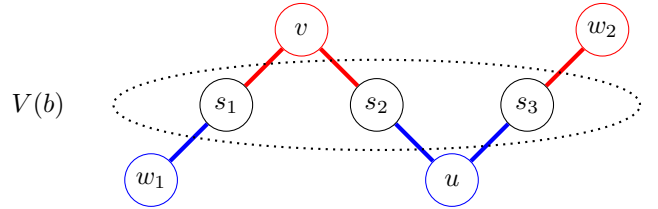
\begin{figure}[b]%
\begin{center}
	\begin{tikzpicture}[auto]
	\node [markovstate] (v1) at (0, 0) {$\sepname_1$};
	\node [markovstate] (v2) at (2, 0) {$\sepname_2$};
	\node [markovstate] (v3) at (4, 0) {$\sepname_3$};
	
	\node [markovstate, draw=blue!200] (v01) at (-1, -1) {$w_1$};
	\node [markovstate, draw=blue!200] (v02) at (3, -1) {$u$};
	
	\node [markovstate, draw=red!200] (v11) at (1, 1) {$v$};
	\node [markovstate, draw=red!200] (v12) at (5, 1) {$w_2$};
	\node (dots1) at (-2.5,0) {$V(\bagname)$};
	%\node (b) at (-2,0) {$V(\bagname)$}

	\draw (v01) [ultra thick, draw=blue!200] to (v1);
	\draw (v2) [ultra thick, draw=blue!200] to (v02);
	\draw (v3) [ultra thick, draw=blue!200] to (v02);
	\draw (v1) [ultra thick, draw=red!200] to (v11);
	\draw (v11) [ultra thick, draw=red!200] to (v2);
	\draw (v3) [ultra thick, draw=red!200] to (v12);
	
		\draw [dotted, thick] (2,0) ellipse (3.5cm and 0.6cm);
	%\draw (v2) to (v3);
	%\draw (v3) to (v4);
	%\draw (v4) [bend left] to (v2);
	%\draw (v4) to (v5);
	%\draw (v4) [bend right, dotted] to (v2);
	%\draw (v4) [bend right, dotted, out=340, in=200] to (v2);
	%\draw (v5) [bend right, dotted] to (v1);
	%\draw (v5) [bend right, dotted, out=320, in=220] to (v1);
	%\draw (v5) [bend right, dotted, out=340, in=200] to (v1);

	%\draw [markovedge] (uv2) to (u1);

	\end{tikzpicture}
\end{center}
\vspace{-0.1cm}
\caption{In this example, the separator is shown in the middle, containing the nodes $V(\bagname)=V(q_1)=V(q_2)=\left\{s_1,s_2,s_3\right\}$. By splitting the path from $w_1$ to $w_2$ along the separator, we obtain multiple paths per side, their number being bounded by the size of the separator. Observe that when two sub-paths, between the same set of waypoints, share a node, this node must be an endpoint for both; otherwise, 
minimality is violated.}%
\label{fig:Join-Split}%
\end{figure}

	\item \textbf{Introduce bags $\bagname$:} Again, we first $1)$ describe the algorithm and then $2)$ prove its correctness. $1)$: For each signature $\sigma_q$ of the child $q$ of $\bagname$, where $V(q) \cup \left\{v\right\}=V(\bagname)$, we first generate all possible combinations of empty paths at $v$. Then, we distribute the link set of $E(\bagname)$ over the endpoints in all possible variations, checking if each distribution can generate some valid signature by possibly moving the endpoints of the subwalks (and possibly, concatenating some). If the answer is yes, we also generate all possible signatures out of these distributions, again by allowing to move the endpoints and allowing to concatenate paths, always respecting capacity constraints. As we only handle $O(1)$ elements, we only perform $O(1)$ operations (covered below). $2)$: Again, assume we did not program some valid signature $\sigma_\bagname$ of $\bagname$. We then obtain a valid signature of $q$ by removing $v$, splitting all paths that traverse it into two, or, if they have $v$ as an endpoint, cutting off $v$, or, if the path only contained $v$, by removing these paths. As the reverse operation will be performed by the prior algorithm, $\sigma_\bagname$ would have been obtained.
\end{itemize}
Each of the above programs be be run in a time of $O(1)$, assuming constant size $\bagname, \tw, \nowp \in O(1)$.
\begin{comment}
In the above programs, we only performed a constant number of operations per bag.
, but we note that individual operations can be costly: If a demand is represented by a binary representation of $x$ bits, each operation can take $O(x)$ time. Hence, the runtime for each program is $O(\log{c_{\max}})$, assuming constant access costs for each node (and their edge connections) in a nice tree decomposition bag.
\end{comment}

Furthermore, we implicitly assumed that for each signature, we also store a representative set of paths s.t.~their total length is minimized. I.e., when generating signatures multiple times for introduce and join nodes, we only keep representatives of minimum total length.
Hence, after dynamically programming the nice tree decomposition $\mathcal T$ bottom-up, 
we consider all solutions at the root node: If an optimal solution exists, it will be represented by a signature, 
and thus, we can choose a walk through the waypoints of minimum length.

It remains to prove the desired runtime of $O(n)$: For constant treewidth $\tw \in O(1)$, we can obtain a nice tree decomposition of width $O(\tw)$ with $O(n)$ bags in a runtime of $O(n)$ using the methods from~\cite{6686186,DBLP:books/sp/Kloks94}. As the dynamic program requires time $O(1)$ for each of the $O(n)$ bags, and as each of the $O(1)$ possible solutions can be checked in time $O(n)$, the claim follows.
\end{IEEEproof}

\subsection{Hardness}

Let us return to the general problem where there can be an arbitrary number of waypoints.
We have shown that for a large graph family of treewidth at most 2,
the outerplanar graphs (which also include cactus graphs for example), 
the routing paths can be computed efficiently. 
This raises the question whether the problem can be solved also
on graphs of treewidth larger than 2, or at least for \emph{all}
graphs of treewidth at most 2. While the latter remains an open question,
in the following we show that problems on graphs of treewidth 3
(namely series-parallel graphs with an additional node connected to all other nodes) are already
NP-hard in general.

\begin{theorem}\label{thm:und-ord-arb-tw3}
The problem of routing through an arbitrary number of waypoints
%Ordered waypoint routing problem is NPC in graphs of treewidth at most~$3$.
is strongly NP-complete on graphs of treewidth at most~$3$.
\end{theorem}
\begin{IEEEproof}
We reduce the ordered waypoint routing problem in graphs of treewidth at most~$3$ 
from the link-disjoint
paths problem in series-parallel graphs, the latter being strongly NP-complete~\cite{Nishizeki2001177}. 

Let~$I$ be an instance of the link-disjoint paths problem in a series
parallel graph~$G$ with terminal pairs~$T_I=\{(s_1,t_1),\ldots,(s_k,t_k)\}$. 
We construct a new instance~$\mathcal{I}$ of the ordered waypoint
problem as follows. Create a graph~$G'\coloneqq G$, then add one new
node~$v$ to~$G'$ and links~$\{t_i,v\},\{s_j,v\}$ for~$i,j\in [k], j\neq 1$,~$i \neq k$.

For simplicity, set for now~$s:=s_1$,~$w_1:=t_1$,~$w_2=v$,~$w_3:=s_2$,~$w_4:=t_2$,
$w_4:=v$,$\dots$,$t:=t_k$, i.e., the order of waypoints is
$s_1,t_1,v,\dots,v,s_i,t_i,v,s_{i+1},t_{i+1},v,\dots,t_k$, with~$3k-2$ waypoints in total.
%%\begin{observation}\label{obsvk-1}
 I.e.,~$v$ ``hosts''~$k-1$ waypoints, with a degree of~$2(k-1)$.
We will show later in the proof 
how to ensure at most one waypoint per node.

\begin{Claim}\label[claim]{clm:countingstar}
In any solution for~$\mathcal{I}$, the union of the~$k-1$
link-disjoint walks from~$s_i$ via~$v$ to~$t_{i+1}$ occupy all links incident to~$v$.
\end{Claim}
\begin{ClaimProof}
Any walk from~$s_i$ via~$v$ to~$t_{i+1}$ must leave and enter~$v$, using
two links. Hence, the union of all these~$k-1$ link-disjoint walks occupy all~$2k-2$ links incident to~$v$.
\end{ClaimProof}

%\begin{observation}\label{obstwplus1}
%\end{observation}
We can now prove the theorem:
%\begin{itemize}
	%\item 
If~$I$ is a yes-instance, then~$\mathcal{I}$ is a
          yes-instance as well: We take the~$k$~$s_i,t_i$-paths from~$I$, connect them in index-order with the~$k-1$ paths
         $t_i,v,s_{i+1}$, and obtain
          the desired ordered waypoint routing.
%\end{itemize}

It is left to show that if~$\mathcal{I}$ is a yes-instance, then~$I$ is a yes-instance as well:
Let~$\mathcal{I}$ be a yes-instance.
Define the path from $s_i$ to $t_i$ as in~$\mathcal{I}$.  As these paths do not use~$v$ or any of the edges adjacent to it
(otherwise the capacity of one of these edges would be exceeded), these paths show that~$I$ is a yes-instance.

On the other hand, the treewidth of~$G'$ is at most the treewidth of~$G$ plus~$1$ (we can
just put~$v$ in all bags of an optimal tree decomposition of~$G$). 
To obtain at most one waypoint on $v$, we create $k-1$ cycles of length four, placing a waypoint on each, and merging another node with $v$. 
This construction does not increase the treewidth and also retains earlier proof arguments.
As
series-parallel graphs have a treewidth of at most 2~\cite[Lemma
11.2.1]{Brandstadt:1999:GCS:302970},~$G'$ has a treewidth of at most~3.
As the problem is clearly in NP, with the reduction being polynomial,
the proof is complete.
\end{IEEEproof}

We conjecture that it is possible to directly modify the proof presented
in~\cite{Nishizeki2001177}, to prove that the feasibility of the waypoint
routing problem is hard even in series-parallel graphs.

\noindent\textbf{One cycle is hard.} In case of non-flow conserving waypoints, NP-hardness strikes earlier already, namely on unicyclic graphs, which contain only one cycle, and as thus have $\tw \leq 2$.
\begin{theorem}\label{thm:unicyclic}
On undirected unicyclic graphs in which waypoints are not flow-conserving, 
computing a route through $O(n)$ waypoints
is weakly NP-complete, even if all waypoints can just increase (or, just decrease) the flow size by at most a constant factor.
\end{theorem}
\begin{IEEEproof}
Reduction from the weakly NP-complete \textsc{Partition} problem~\cite{DBLP:books/fm/GareyJ79}, where an instance $I$ contains $\ell$ non-negative integers $i_1,\dots, i_{\ell}$, $\sum_{j=1}^{\ell}i_j=S$, with the size of the binary representation of all integers polynomially bounded in $\ell$.

We begin with the case that waypoints can change the flow size arbitrarily.
W.l.o.g., let $\ell$ be even and $i_1 \leq i_2 \leq \dots \leq i_{\ell}$. We create two stars (denoted left and right star) with $1+\ell/2$ leaf nodes each, where all links have a capacity of $S$.
We connect both star center nodes in a cycle
, with the cycle links having a capacity of $S/2$ each, respectively.

Next, we place $s$, here also identified as $w_1$, on a leaf of the left star and $t$ on a leaf in the right star.
To distribute the remaining $\ell-1$ waypoints $w_2,\dots,w_{\ell}$, corresponding to the integers, we place the ones with even indices on leaves in the left star, and those with odd indices in the right star.

Suppose the routing starts with a size of $i_1$, is changed to $i_2$ by $w_2$ and so on. Then, solving the \textsc{Partition} instance $I$ is equivalent to computing a waypoint routing, as the paths going along the cycle have to be partitioned into two sets, each having a combined demand of $S/2$. 

So far, we assumed that waypoints can change the flow size arbitrarily -- but hardness also holds if each waypoint can just increase (or, just decrease) the flow size by a constant amount.
In order to do so, we replace the leaf nodes of the stars with paths of $O(\log S)$ waypoints, which are used to increase the demands to the desired size.
\end{IEEEproof}

The directed graph case is analogous by putting all waypoints to one star, creating the same amount of intermediate dummy waypoints in the other star, which do not change the flow size, and replacing all undirected links with two directed links of opposite directions and identical capacity.

\begin{corollary}\label{corr:unicyclic-cactus}
On directed graphs, with the underlying undirected graph being unicyclic and where waypoints are not flow-conserving, 
computing a route through $O(n)$ waypoints
is NP-complete, even if all waypoints can just increase (or, just decrease) the flow size by at most a constant factor.
\end{corollary}

For these two proofs, we used flow sizes that can be exponential in the graph size (binary encoded).
Nonetheless, recall \Cref{thm:und-cap,thm:hard-directed}, where we showed that the problem also stays strongly NP-complete on general graphs.

\section{Other Related Work}\label{sec:relwork}

In this paper, we focus on the allocation of a \emph{single}
walk, without violating capacity constraints. 
However, there also exists literature on how to \emph{admit} and allocate 
\emph{multiple} 
walks, e.g., using randomized rounding and tolerating some capacity augmentation~\cite{sss-guy,sirocco15,sirocco16path}; there are also extensions to more complex requests
such as trees~\cite{sirocco16path, bansal2011minimum}.

Moreover, while we focused on  walks through \emph{ordered} waypoints,
there is work on routing through \emph{unordered} waypoints~\cite{Foerster2017,2017arXivunordered}.
The problem of finding shortest (link- and node-disjoint)
paths and cycles through a \emph{set} of $k$ waypoints has
been a central topic of graph theory for several decades~\cite{DBLP:journals/ijfcs/PerkovicR00}.
A cycle from $s$ through $k=1$
waypoints back to $t=s$ can be found efficiently by
breadth first search, for $k=2$ the problem 
corresponds to finding a integer
flow of size 2 between
two nodes, and for $k = 3$, it can still be solved in linear time~\cite{DBLP:journals/ipl/FleischnerW92,DBLP:journals/tcs/FortuneHW80};
a polynomial-time solution for any constant $k$ follows from the work 
on the disjoint path problem~\cite{DBLP:journals/jct/RobertsonS95b}. 
The best known deterministic algorithm to compute \emph{feasible} 
(but not necessarily shortest) paths is by 
Kawarabayashi~\cite{DBLP:conf/ipco/Kawarabayashi08}: it finds a cycle for up to $k = O((\log \log n)^{1/10})$
waypoints in deterministic polynomial time. 
Bj\"orklund et al.~\cite{thore-soda} presented a randomized algorithm 
based on algebraic techniques
which finds a shortest simple cycle through a given set of $k$ node or links in an 
$n$-node undirected graph in time $2^kn^{O(1)}$.
However, there is no obvious way to apply algorithms designed for the unordered problem variant 
to the 
ordered
 problem 
variant.

\section{Conclusion}\label{sec:conclusion}

This paper initiated the study of a fundamental
algorithmic problem underlying modern network
services: the routing via waypoints using walks. 
We hope that our paper can 
provide
  the network community
with algorithmic techniques but also inform about 
complexity bounds.
While we present a comprehensive set of algorithms
and hardness results, there are several interesting
directions for future research, for example regarding randomized
algorithms or algorithms for scenarios where capacity
constraints may be violated slightly.

\vspace{0.1cm}
\noindent \textbf{Acknowledgments.}
We like to thank Thore Husfeldt for inspiring discussions.
Research partly supported by the Villum project ReNet as well as by Aalborg University's PreLytics project.
Saeed Amiri's research was partly supported by the European Research
Council (ERC) under the European Union’s Horizon 2020 research and innovation programme
(grant agreement No 648527).
%\end{comment}

\balance

%\clearpage
%\bibliographystyle{IEEEtran}
%\bibliography{literature}

\end{document}